\documentclass[twocolumn]{aastex631}

\usepackage {amsmath}
\usepackage{multirow}
\received{}
\revised{}
\accepted{}
\submitjournal{ApJ}

\shorttitle{Erosion and accretion by cratering impacts on rocky and icy bodies}
\shortauthors{R. Hyodo \& H. Genda}

\graphicspath{{./figs/}}

\begin{document}
\title{Erosion and accretion by cratering impacts on rocky and icy bodies}

\correspondingauthor{Ryuki Hyodo}
\email{ryuki.h0525@gmail.com}
\author[0000-0003-4590-0988]{Ryuki Hyodo}
\affiliation{ISAS, JAXA, Sagamihara, Japan}

\author[0000-0001-6702-0872]{Hidenori Genda}
\affiliation{Earth-Life Science Institute, Tokyo Institute of Technology, Tokyo, 152-8550, Japan}

\begin{abstract}
During planet formation, numerous small impacting bodies result in cratering impacts on large target bodies. A fraction of the target surface is eroded, while a fraction of the impactor material accretes onto the surface. These fractions depend upon the impact velocities, the impact angles, and the escape velocities of the target. This study uses smoothed particle hydrodynamics simulations to model cratering impacts onto a planar icy target for which gravity is the dominant force and material strength is neglected. By evaluating numerical results, scaling laws are derived for the escape mass of the target material and the accretion mass of the impactor material onto the target surface. Together with recently derived results for rocky bodies in a companion study, a conclusion is formulated that typical cratering impacts on terrestrial planets, except for those on Mercury, led to a net accretion, while those on the moons of giant planets, e.g., Rhea and Europa, led to a net erosion. Our newly derived scaling laws would be useful for predicting the erosion of the target body and the accretion of the impactor for a variety of cratering impacts that would occur on large rocky and icy planetary bodies during planet formation and collisional evolution from ancient times to today.
\end{abstract}


%%%%%%%%%%%%%%%%%%%%%%%%%
% Introduction
%%%%%%%%%%%%%%%%%%%%%%%%%
\section{Introduction} \label{sec:intro}

Cratering impacts -- collisions of small impacting bodies on large planetary bodies -- erode a portion of the large planetary surface by ejecting materials with a speed greater than the escape velocity for the large planetary body. Simultaneously a fraction of impactor's material is accreted onto the target surface \citep[e.g.,][and references therein]{Hou83,Sve11,Hou11,Hyo20}. The erosion of the target's surface materials and the accretion of the impactor's materials are key processes that dictate the mass gain/loss of planetary bodies and that characterize the surface composition \citep[e.g.,][]{Mel84,Mel89}. 

Using a point-source model, the crater size is assumed to be much larger than that of the impacting body, an analytical scaling law that predicts the amount of the ejecta $m_{\rm eje}(>v_{\rm eje})$ whose ejection velocity is larger than a given value $v_{\rm eje}$ is derived \citep[e.g.,][]{Hou11,Hol12}. The amount of the target material that is eroded by escaping from its own gravity upon cratering impacts (hereafter, the escape mass $m_{\rm esc,tar}$) increases with increasing impact velocity $v_{\rm imp}$ compared to the escape velocity of the target $v_{\rm esc}$. 

In our companion study \citep[][hereafter Paper 1]{Hyo20}, the fraction of escape material from the target body $m_{\rm esc,tar}$ was studied for a variety of cratering impacts for {\it rocky} bodies. Using smoothed particle hydrodynamics (SPH) simulations, Paper 1 demonstrated that the point-source scaling law agreed with the numerically derived escape mass of the target when $v_{\rm imp} \gtrsim 12 v_{\rm esc}$. Otherwise, it was found to overestimate the escape mass when the ejection velocity of the target material is only moderately larger than the escape velocity of the target.

The escape mass as well as the applicability of the point-source scaling law have not been studied for the case of {\it icy} bodies. In the outer solar system, icy bodies are ubiquitous. Following the same approach as Paper 1, an extensive number of cratering impact simulations onto a planar {\it icy} target were performed. These simulations consider a case where gravity dominates and the material strength may be neglected. We independently study the high-speed escaping ejecta (i.e., $v_{\rm eje} > v_{\rm esc}$) that originate from the icy target or the icy impactor. A goal of this study is to derive scaling laws for the escape mass of the target material $m_{\rm tar,esc}$ and that of impactor $m_{\rm imp,esc}$. Considering the mass balance, the mass originating from the impactor that are deposited onto the target surface upon cratering is given as $m_{\rm imp,acc}$ as $m_{\rm imp,acc} = m_{\rm imp} - m_{\rm imp,esc}$ ($m_{\rm imp}$ is the impactor mass).

The motivation of this work is to provide scaling laws that would be useful for the planetary community. These scaling laws would better predict the fraction of the eroded target body materials which escape the gravity and the fraction of impactor materials deposited on the target surface during a cratering impact in the gravity-dominated regime. The scaling laws are provided as a function of the impact velocity $v_{\rm imp}$, the impact angle $\theta$, and the escape velocity of the target $v_{\rm esc}$ (see also Paper 1 for rocky bodies).

The numerical methods are presented in section \ref{sec_method}. In section \ref{sec_target}, the numerical results of the ejecta that originate from the target (the target material) are presented. Section \ref{sec_target} also presents the derivations of new scaling laws for the escape mass of the target material. In section \ref{sec_impactor}, the numerical results of the ejecta originating from the impactor (the impactor material) are presented. A new scaling law for the accretion mass of the impactor material onto the target surface is also presented. In Section \ref{sec_application}, a discussion of applications using the newly derived scaling laws for planet formation is presented. Finally, in section \ref{sec_summary} a summary is presented.

%%%%%%%%%%%%%%%%%%%%%%%%%%%%%%%%%%%%%%%%
% Section 2
%%%%%%%%%%%%%%%%%%%%%%%%%%%%%%%%%%%%%%%%
\section{Numerical methods} \label{sec_method}

The fate of high-speed ejecta with exceeding the escape velocity of the {\it icy} target bodies for a given impact velocity and impact angle were evaluated. This was performed using 3D SPH simulations \citep{Luc77,Mon92} for cratering impacts (see also Paper 1 for rocky bodies). The relevant equations of state (EOSs) was a 5-Phase EOS for icy materials \citep{Sen08}. This 5-Phase EOS included the effects of melting and vaporization for H$_{2}$O. The numerical code was identical as that used to study cratering impacts \citep{Kur19,Hyo19} and as that used to study a catastrophic disruption of icy bodies \citep{Hyo17}. In this work, high-speed ejecta whose ejection speed exceeds the escape velocity was considered. Material strength were neglected in the simulations. 

The icy impactor was idealized as a spherical projectile with a radius of $R_{\rm imp}=10$ km. The icy target was idealized as a flat surface contained in a half-sphere with a radius of typically greater than that of the impactor (projectile) by a factor of 10-20. For the numerical resolution, the total number of equal-mass SPH particles (for the target and for the impactor) used in our simulations was $\sim 1.7 \times 10^7$. As a resolution test, the total number of particles was kept constant but kept changing the target size and the number of particles to resolve impactors. $\sim 3.4 \times 10^4$ and $\sim 4.8 \times 10^3$ SPH particles were used for the impactors which correspond to $20$ and $10$ SPH particles per projectile radius (PPPR), respectively. Convergence of the numerical results was confirmed by comparing 10 and 20 PPPR cases for the total mass of the high-speed ejecta of $v_{\rm eje} > v_{\rm esc}$ (with $v_{\rm esc} = 1-10$ km s$^{-1}$).

Different sources of the escape mass, the target material (Section \ref{sec_target}) and impactor material (Section \ref{sec_impactor}), were independently investigated. Ejecta particles were defined as those existing above the surface of the target for the analysis. Snapshots were utilized when the shock reached the boundary of the target for the analysis. 

Various impact velocities typically ranging from $6 - 62$ km s$^{-1}$ (7 km s$^{-1}$ interval) and at various impact angles from $15 - 90$ degrees ($15$ degrees interval), where a 90-degree impact is a head-on/vertical impact, were analyzed. In dimensionless form, model fidelity increases as the ratio of the large planetary body size to that of the cratering impacts increases such that the curvature of the target is of negligible importance \citep{Gen17}. The results can be related to any impactor size because all hydrodynamic equations can be rewritten in a dimensionless form without gravity and strength \citep{Joh12}.

%%%%%%%%%%%%%%%%%%%%
% Figure 1 with double column
%%%%%%%%%%%%%%%%%%%%
\begin{figure*}
	\centering
	  \includegraphics[width=1\textwidth]{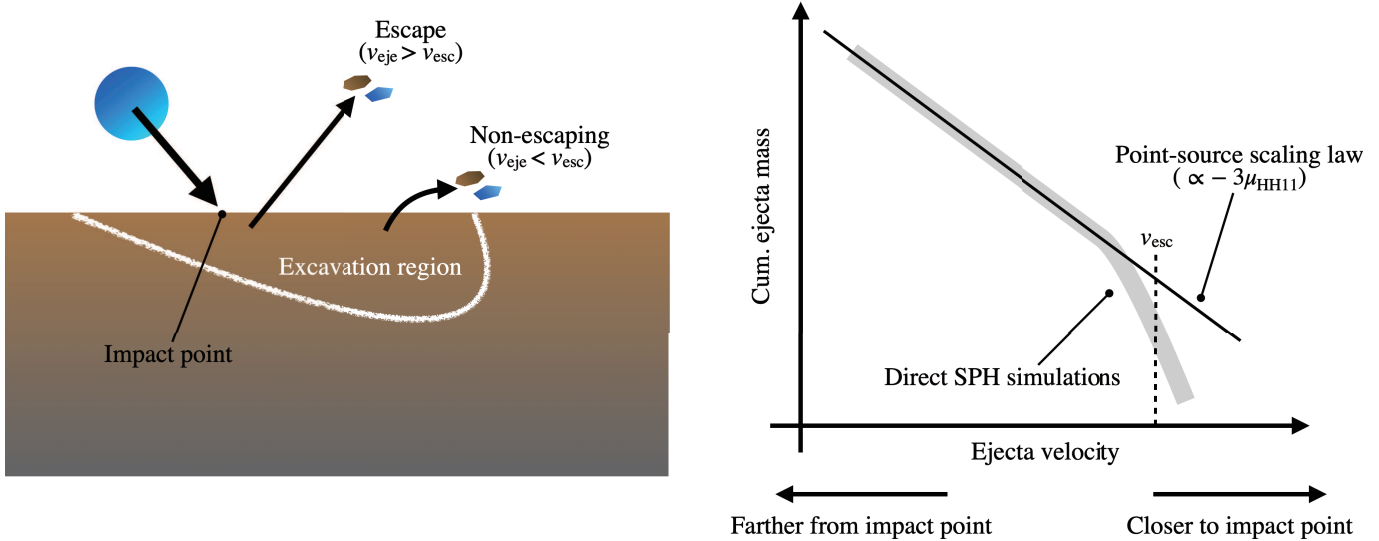}
	\caption{Schematic illustration of cratering impacts (left) and velocity distribution of the ejecta (right). High- and low-speed ejecta are launched closer to and farther from the impact point, respectively; that is, the escape mass is launched from the vicinity of the impact point (left). Irregular-shaped objects in the left panel indicate ejecta of target and impactor materials (the total amount and their fractions depend on the impact condition). The point-source scaling law is characterized by a unique slope $-3\mu_{\rm HH11}$ in the velocity distribution (the black line in the right panel; Equation \ref{eq_HH11}), while direct SPH simulations of cratering impacts of icy bodies demonstrated that high-speed ejecta would have a steeper slope for the velocity distribution (the gray line in the right panel; see also Paper 1). This discrepancy leads to an overestimation in the escape mass when using the point-source scaling law, depending on the considered escape velocity of the target $v_{\rm esc}$ (the same was reported for rocky bodies; Paper 1).}
\label{fig_schematic}
\end{figure*}

%%%%%%%%%%%%%%%%%%%%%%%%%%%%%%%%%%%%%%%%
% Section 3
%%%%%%%%%%%%%%%%%%%%%%%%%%%%%%%%%%%%%%%%
\section{Escape of target material} \label{sec_target}

In this section, the escape mass that originates from the target (the target material) is discussed. First, the point-source scaling law (Section \ref{sec_HH11}) is explained. In Section \ref{sec_ejection}, the ejection velocity distributions for different impact angles are investigated. In Section \ref{sec_scaling_target}, a new scaling law for the escape mass of the target materials is derived. Finally, the newly derived scaling law is compared to the classical point-source scaling law (Section \ref{sec_comparison}).

%%%%%%%%%%%%%%%%%%%%%%%%%%%%%%%%%%%%%%%%%%%%%%%%%%%%
\subsection{The point-source scaling law} \label{sec_HH11}

Cratering impacts lead to an ejection of the target material from the surface. Pioneering studies \citep{Hou83,Hol87,Hol93,Hol07,Hou11} predicted $m(>v_{\rm eje,tar})$ as a function of the impact velocity $v_{\rm imp}$ and impact angle $\theta$ assuming that the size of the impact crater is assumed to be much larger than that of the impactor (hereafter $M_{\rm HH11,esc,tar}$)\footnote{We note that the point-source scaling law implicitly assumed negligible contribution from the impactor material.}. The point-source scaling law can potentially predict the escape mass by using the relation, $v_{\rm eje,tar}=v_{\rm esc}$, where $v_{\rm esc}$ is the escape velocity of the target, as follows (hereafter HH11):

%%%%%%%%%%%
% Equation 1
%%%%%%%%%%%
\begin{equation}
\label{eq_HH11}
	\frac{M_{\rm HH11,esc,tar}(>v_{\rm esc})}{m_{\rm imp}} = C_{\rm HH11} \left( \frac{v_{\rm esc}}{v_{\rm imp}\sin \theta} \right)^{-3\mu_{\rm HH11}} ,
\end{equation}

\noindent where $C_{\rm HH11}=(3k/4\pi)C_{0}^{3\mu_{\rm HH11}}$ and $\mu_{\rm HH11}$ are constants. From the experiments, $C_{\rm 0}=1.5$ and $\mu_{\rm HH11}=0.55$ for nonporous icy materials and rocky materials, respectively \citep[e.g.,][]{Hol07}. $k=0.2$ for ice and $k=0.3$ for rock, respectively \citep{Gau63}, respectively. The densities of the impactor and target are assumed to be identical.

%%%%%%%%%%%%%%%%%%%%%%%%%%%%%%%%%%%%%%%%%%%%%%%%%%%%
\subsection{Velocity distribution of the ejecta} \label{sec_ejection}

%%%%%%%%%%%%%%%%%%%%
% Figure 2 with double column
%%%%%%%%%%%%%%%%%%%%
\begin{figure*}[ht!]
 \centering \includegraphics[width=\textwidth]{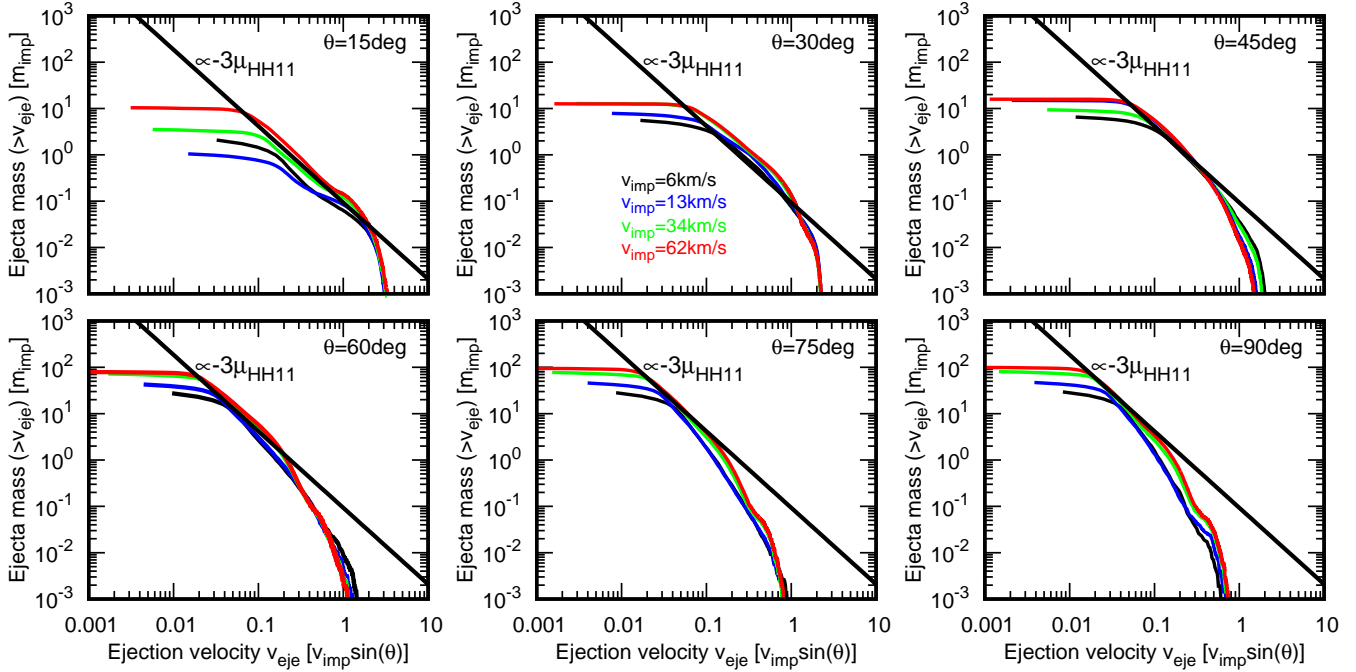}
\caption{Cumulative ejecta mass ($>v_{\rm eje}$; target material) that exceeds a given ejection velocity $v_{\rm eje}$ for a different impact angle $\theta$. The black, blue, green, and red lines represent the results of SPH simulations for $v_{\rm imp}=6, 13, 34$, and $62$ km s$^{-1}$, respectively. The black line represents the point-source scaling law ($-3\mu_{\rm HH11}$; Equation \ref{eq_HH11}). Numerical results do not converge for small velocities where the point-source scaling assumption is valid, but this does not affect the conclusion of the current study (since our primary interest is the high-speed ejecta of $v_{\rm eje} > v_{\rm esc}$).}
\label{fig_velocity}
\end{figure*}

The impact shock propagates through the interior of the target after the impactor makes contact with the target surface. The crater forms as the target material is sheared, moving upward and outward along the bowl-shaped crater edge \citep[Figure \ref{fig_schematic}; see also][]{Hou11}. The speed of the ejecta $v_{\rm eje}$ depends on its launch position \citep[e.g.,][]{Pie80}, that is, high- and low-speed ejecta are launched closer to and farther from the impact point, respectively (Figure \ref{fig_schematic}).

Figure \ref{fig_velocity} shows the cumulative mass of ejecta originating from the target as a function of ejection velocity $v_{\rm eje}$ for different impact velocities and impact angles. The ejection velocity is scaled by $v_{\rm imp} \sin \theta$. Note that the data for the ejection velocity with $v_{\rm eje} \lesssim 0.1 v_{\rm imp} \sin \theta$ failed to converge during the performed simulation time (i.e., more numerical time is needed for the convergence of smaller $v_{\rm eje}$ because the lower-speed ejecta is launched farther from the impact point\footnote{It was noted that the point-source scaling law would not be appropriate for a very small ejection velocity because of the effects of material strength and gravity \citep[e.g.,][]{Hou11}, although such a regime does not pertain to this study.}). The point-source scaling law (Equation \ref{eq_HH11}) is plotted as a black line. The results indicate that the distribution is uniquely scaled by $v_{\rm imp} \sin \theta$, which is in accordance with the basic principles evident in the point-source assumption \citep{Hol07}.

The slope of the ejection velocity distribution is characterized by $-3\mu_{\rm HH11}$, a unique constant value, in the case of the point-source scaling law (Equation \ref{eq_HH11}). This agrees with the results obtained from direct impact simulations only for the low-speed ejecta (Figure \ref{fig_velocity}). This indicates that the point-source scaling law is valid only for a limited range of ejection velocity distributions.

The ejecta velocity distribution has a steeper slope for high-speed ejecta than that predicted by the point-source scaling law (see also Figure \ref{fig_schematic}). The ejection velocity distribution for such high-speed ejecta has a unique power-law dependence depending on different impact angles (Figure \ref{fig_velocity}). This result indicates that the point-source scaling law overestimates the mass of the high-speed ejecta with a launch point near the impact point. This was also reported in the case of rocky bodies (Paper 1). In the next subsection, updated scaling laws are derived for the high-speed ejecta where the point-source scaling law is not applicable.

%%%%%%%%%%%%%%%%%%%%%%%%%%%%%%%%%%%%%%%%%%%%%%%%%%%%
\subsection{Updated scaling law for high-speed ejecta} \label{sec_scaling_target}

%%%%%%%%%%%%%%%%%%%%
% Figure 3 with double column
%%%%%%%%%%%%%%%%%%%%
\begin{figure*}[ht!]
\centering \includegraphics[width=0.75\textwidth]{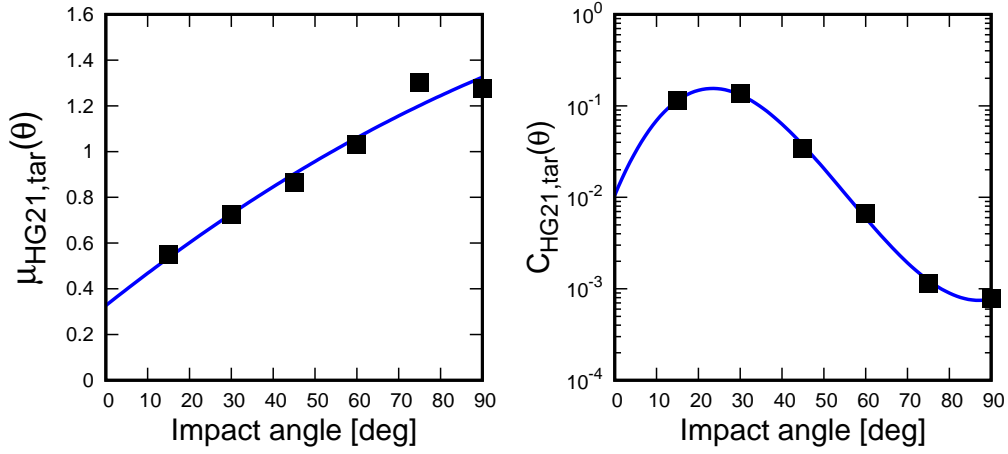}
\caption{The exponent $\mu_{\rm HG21,tar}(\theta)$ (left) and the coefficient $C_{\rm HG21,tar}(\theta)$ (right) for the new scaling law of the escape mass of the target material (Equations \ref{eq_HG21_target} and \ref{eq_new_target}) as a function of the impact angle. The points represent the results of SPH simulations and solid curves are defined by fitted quadratic and cubic functions of the impact angle for $\mu_{\rm HG21,tar}(\theta)$ and $C_{\rm HG21,tar}(\theta)$, respectively.}
\label{fig_mu_C_target}
\end{figure*}

%%%%%%%%%%%%%%%%%%%%
% Figure 4 with double column
%%%%%%%%%%%%%%%%%%%%
\begin{figure*}[ht!]
\centering \includegraphics[width=\textwidth]{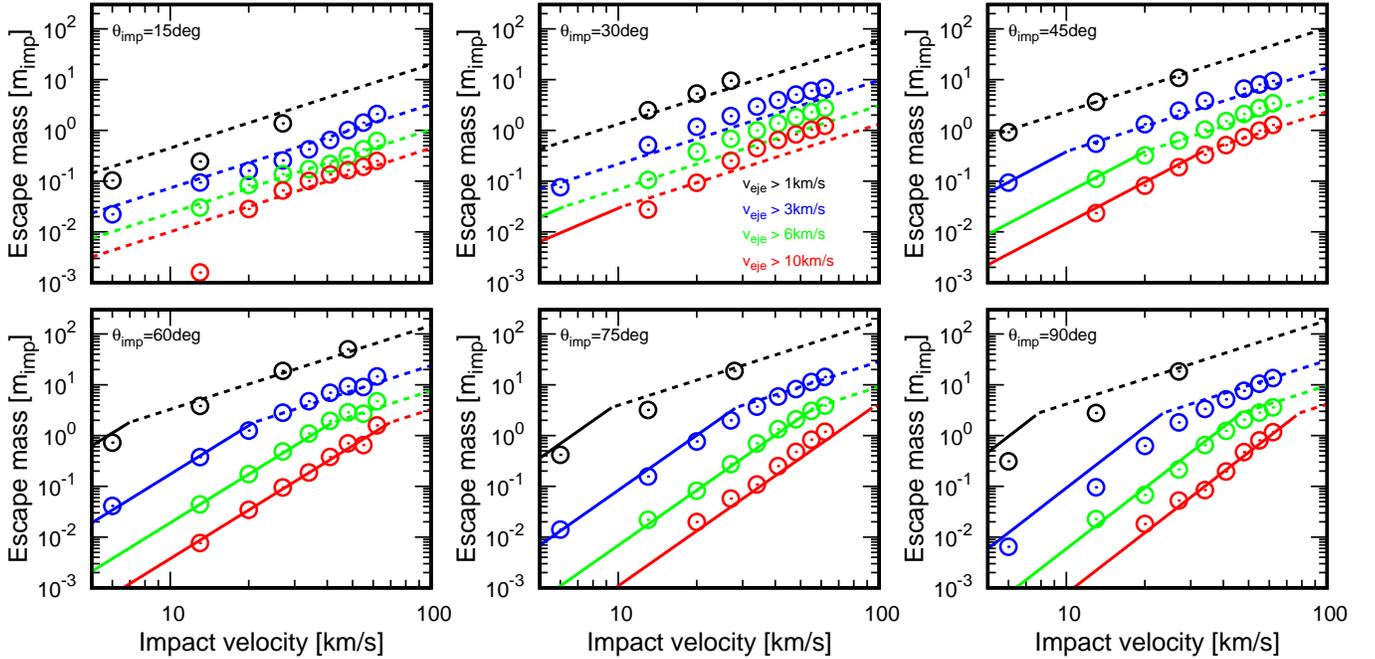}
\caption{Escape mass of target material (in units of $m_{\rm imp}$) as a function of $v_{\rm imp}$ for different $\theta$. Points are the results of the SPH simulations. Solid and dashed lines represent the updated scaling law (Equation \ref{eq_new_target}; $\min \left\{ M^{*}_{\rm HG21,esc,tar}, M_{\rm HH11,esc,tar} \right \}$). Solid lines represent the cases where the same function as in Equation \ref{eq_HG21_target} ($M^{*}_{\rm HG21,esc,tar}$) are used. Dashed lines depict cases in which the function is the same as in Equation \ref{eq_HH11} ($M_{\rm HH11,esc,tar}$). The black, blue, green, and red lines represent cases where $v_{\rm esc}=1, 3, 6$, and $10$ km s$^{-1}$, respectively.}
\label{fig_target}
\end{figure*}

Following the methodology of Paper 1 and from the arguments in the previous subsection, the escape mass of the target material $M^{*}_{\rm HG21,esc,tar}$ for the high-speed ejecta regime ($\gtrsim 0.1 v_{\rm imp} \sin \theta$) is uniquely expressed by a power-law as a function of impact velocity and impact angle as follows:

%%%%%%%%%%%
% Equation 3
%%%%%%%%%%%
\begin{align}
\label{eq_HG21_target}
	\frac{M^{*}_{\rm HG21,esc,tar}(>v_{\rm esc})}{m_{\rm imp}} = \nonumber \\ 
	 C_{\rm HG21,tar}(\theta) & \left( \frac{v_{\rm esc}}{v_{\rm imp}\sin \theta } \right)^{-3\mu_{\rm HG21,tar}(\theta)} ,
\end{align}

\noindent where $C_{\rm HG21,tar}(\theta)$ and $\mu_{\rm HG21,tar}(\theta)$ are a new coefficient and a new exponent that depends on the impact angle, respectively. 

One may obtain the exact $C_{\rm HG21,tar}(\theta)$ and $\mu_{\rm HG21,tar}(\theta)$ at varying impact angles using the numerical results (indicated by blue points in Figure \ref{fig_mu_C_target}). Subsequently, the approximated $\mu_{\rm HG21,tar}(\theta)$ and $C_{\rm HG21,tar}(\theta)$ (the black lines in Figure \ref{fig_mu_C_target}) were derived using the quadratic and cubic functions of the impact angle (the black lines in Figure \ref{fig_mu_C_target}) as

%%%%%%%%%%%
% Equation 3&4
%%%%%%%%%%%
\begin{align}
\label{eq_mu_HG21}
	& \mu_{\rm HG21,tar}(\theta) = a_{\rm tar}\theta^2 + b_{\rm tar}\theta + c_{\rm tar} \\
\label{eq_C_HG21}
	& C_{\rm HG21,tar}(\theta) = \exp \left( d_{\rm tar}\theta^3 + e_{\rm tar}\theta^2 + f_{\rm tar}\theta + g_{\rm tar} \right) ,
\end{align}

\noindent where $a_{\rm tar}$, $b_{\rm tar}$, $c_{\rm tar}$, $d_{\rm tar}$, $e_{\rm tar}$, $f_{\rm tar}$, and $g_{\rm tar}$ are the fitted parameters, respectively (Table \ref{table_param}).

Figure \ref{fig_target} shows the escape mass of the target material (target mass with an ejection velocity greater than the escape velocity) as a function of the impact velocity. Points were obtained from the numerical simulations. 

As Paper 1 established that for rocky bodies, the original point-source scaling law for icy bodies (HH11; Equation \ref{eq_HH11}) is valid only at sufficiently large distances from the impact point (i.e., for a sufficiently small ejection velocity). The coefficient and exponent deviate from those of HH11 (compare the solid black line and the other lines in Figure \ref{fig_velocity}) for large ejection velocities (i.e., small distances to the impact point). 

As observed in Figure \ref{fig_velocity} (see also Figure \ref{fig_schematic}), HH11 for icy bodies overestimates the escape mass of the target material, especially for a significant ejection velocity. Therefore, the new scaling law for the escape mass of the target material that combines with HH11 is given by $\min \left\{ M^{*}_{\rm HG21,esc,tar}, M_{\rm HH11,esc,tar} \right \}$ and is written as follows (hereafter, HG21):

%%%%%%%%%%%
% Equation 5
%%%%%%%%%%%
\begin{align}
\label{eq_new_target}
	 & \frac{M_{\rm HG21,esc,tar}(>v_{\rm esc})}{m_{\rm imp}} = \nonumber \\
	 & \min \left\{  C_{\rm HG21,tar}(\theta) \left( \frac{v_{\rm esc}}{v_{\rm imp}\sin(\theta)} \right)^{-3\mu_{\rm HG21,tar}(\theta)}, \right. \nonumber \\ 
	 & \hspace{2.5cm} \left. C_{\rm HH11} \left( \frac{v_{\rm esc}}{v_{\rm imp}\sin(\theta)} \right)^{-3\mu_{\rm HH11}}  \right\} . 
\end{align}

The new scaling law (Equation \ref{eq_new_target}) is plotted by solid and dashed lines in Figure \ref{fig_target}. A close match between the updated scaling law and impact simulations (points) for $v_{\rm imp} = 6 - 62 $ km s$^{-1}$ and $\theta = 15 - 90 $ degrees for $v_{\rm esc}= 1 - 10 $ km s$^{-1}$ was observed.

%%%%%%%%%%%%%%%%%%%%
% Table
%%%%%%%%%%%%%%%%%%%%
\begin{table}[]
\begin{center}
\begin{tabular}{|c|c|c|c|}
\hline

$a_{\rm tar}$         & $b_{\rm tar}$          & $c_{\rm tar}$         &                 \\ \hline
$-3.76 \times 10^{-5}$ & $1.45 \times 10^{-2}$  & $3.27 \times 10^{-1}$ &  \\ \hline
$d_{\rm tar}$         & $e_{\rm tar}$          & $f_{\rm tar}$         & $g_{\rm tar}$ \\ \hline
$4.13 \times 10^{-5}$ & $-6.85 \times 10^{-3}$ & $2.53 \times 10^{-1}$ & $-4.55$ \\ \hline

\hline
\hline

$a_{\rm imp}$         & $b_{\rm imp}$          & $c_{\rm imp}$         &                      \\ \hline
$-3.61 \times 10^{-4}$ & $4.48 \times 10^{-2}$  & $-5.99 \times 10^{-1}$ &          \\ \hline
$d_{\rm imp}$         & $e_{\rm imp}$          & $f_{\rm imp}$         & $g_{\rm imp}$ \\ \hline
$3.97\times 10^{-5}$ & $-4.74 \times 10^{-3}$ & $4.99 \times 10^{-2}$ & $2.53 \times 10^{-1}$ \\ \hline

\end{tabular}

\caption{Parameters of the fitted polynomial and exponential functions for the escape masses of the target material (Equation \ref{eq_new_target}) and the impactor material (Equation \ref{eq_HG21_impactor}), respectively. Those for rocky bodies are shown in Paper 1.}
\label{table_param}
\end{center}
\end{table}

%%%%%%%%%%%%%%%%%%%%%%%%%%%%%%%%%%%%%%%%%%%%%%%%%%%%
\subsection{Comparison to the point-source scaling law} \label{sec_comparison}

The point-source scaling law (Equation \ref{eq_HH11}) has been widely used. As demonstrated in the case of the rocky bodies (Paper 1), the point-source scaling law for the escape mass of the target material of icy bodies also agrees with the numerical results when $v_{\rm imp} \gg v_{\rm esc}$, whereas it overestimates the escape mass when $v_{\rm imp} \gtrsim v_{\rm esc}$. 

\begin{figure}[ht!]
	\centering
\includegraphics[width=0.4\textwidth]{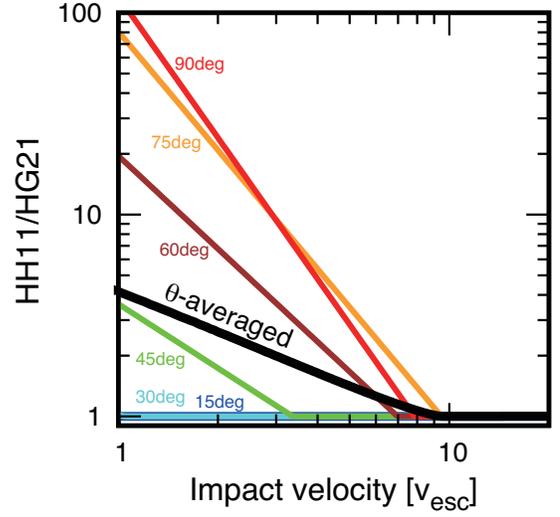}
\caption{The ratio of the point-source scaling law (HH11; Equation \ref{eq_HH11}) to the newly derived scaling law (HG21; Equation \ref{eq_new_target}) for the escape target mass of icy bodies. The $x$-axis represents the impact velocity in a unit of escape velocity. The red, orange, brown, green, cyan, and blue lines are the cases of the impact angles of $\theta=$90, 75, 60, 45, 30, and 15 degrees, respectively. The solid black line represents the $\theta$-averaged case over $\sin 2\theta$. Those for rocky bodies are shown in Paper 1.}
\label{fig_comparison}
\end{figure}

Figure \ref{fig_comparison} shows the overestimation of the point-source scaling law (HH11; Equation \ref{eq_HH11}) when compared to the results of the direct impact simulations (HG21; Equation \ref{eq_new_target}) as a function of the impact velocity. Except for impact angles of $15$ and $30$ degrees, HH11 overestimated the escape mass for $v_{\rm imp} \lesssim 9 v_{\rm esc}$, which exponentially increased as the impact velocity decreased. As found in the case of the rocky bodies (Paper 1), the difference became more significant as the impact angle tended toward a head-on collision. HH11 overestimated the escape mass of the target material by approximately $\sim 100$ times when $v_{\rm imp}/v_{\rm esc} \sim 1$ at vertical impact ($\theta=90$ degrees). For the $\theta$-averaged escape mass over the $\sin 2\theta$ distribution (Equations \ref{eq_theta_average_target_ice1}--\ref{eq_theta_average_target_ice2}), HH11 overestimated by a factor of $\sim 4$ larger than HG21 when $v_{\rm imp}/v_{\rm esc} \sim 1$.

%%%%%%%%%%%%%%%%%%%%
% Figure 5 with single column
%%%%%%%%%%%%%%%%%%%%

%%%%%%%%%%%%%%%%%%%%%%%%%%%%%%%%%%%%%%%%
% Section 4
%%%%%%%%%%%%%%%%%%%%%%%%%%%%%%%%%%%%%%%%
\section{Accretion of impactor material} \label{sec_impactor}

In this section, a discussion regarding the accretion mass of the impactor material onto the target surface upon cratering impacts is presented. As a consequence of a cratering impact, a fraction of the impactor material (mass of $m_{\rm imp,esc}$) escapes as its speed exceeds the escape velocity of the target, and the remaining material accretes onto the target surface. 

The accretion mass originating from the impactor, the accretion mass of the impactor material $m_{\rm imp,acc}$, is given by the mass balance as $m_{\rm imp,acc} = m_{\rm imp} - m_{\rm imp,esc}$. By evaluating $m_{\rm imp,esc}$ using SPH simulations and by considering the mass balance, one obtains $m_{\rm imp,acc}$. A scaling law of the accretion mass of the impactor material onto the target surface is derived, as shown below.

%%%%%%%%%%%%%%%%%%%%
% Figure 6 with double column
%%%%%%%%%%%%%%%%%%%%
\begin{figure*}
\centering \includegraphics[width=0.75\textwidth]{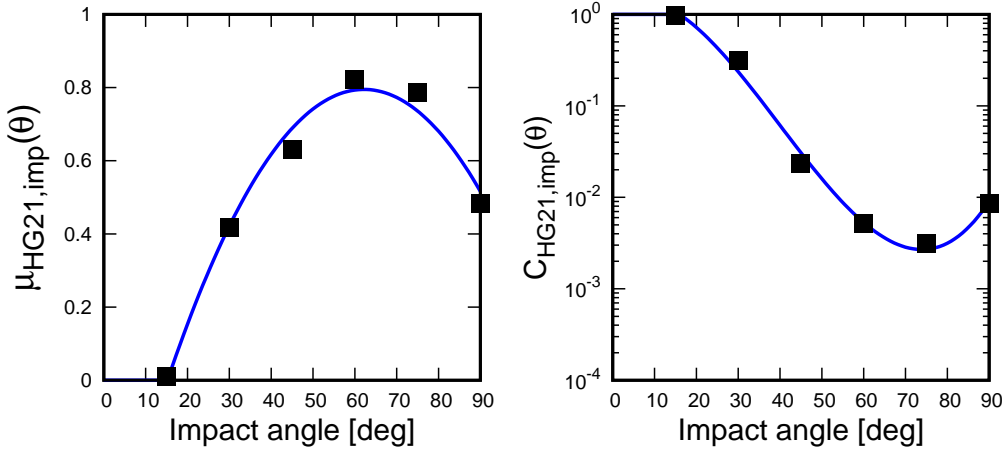}
	\caption{The exponent $\mu_{\rm HG21,imp}(\theta)$ (left) and the coefficient $C_{\rm HG21,imp}(\theta)$ (right) for the new scaling law (Equation \ref{eq_HG21_impactor}) that predicts the escape mass of the impactor material as a function of the impact angle. The points are the results of SPH simulations and solid curves are the fitted quadratic and cubic functions of the impact angle for $\mu_{\rm HG21,imp}(\theta)$ and $C_{\rm HG21,imp}(\theta)$, respectively. Note that $\mu_{\rm HG21,imp}(\theta) = 0$ and $C_{\rm HG21,imp}(\theta) = 1$ for $\theta < 15$ degrees.}
	\label{fig_mu_C_impactor}
\end{figure*}

%%%%%%%%%%%%%%%%%%%%
% Figure 7 with double column
%%%%%%%%%%%%%%%%%%%%
\begin{figure*}[ht!]
\centering \includegraphics[width=\textwidth]{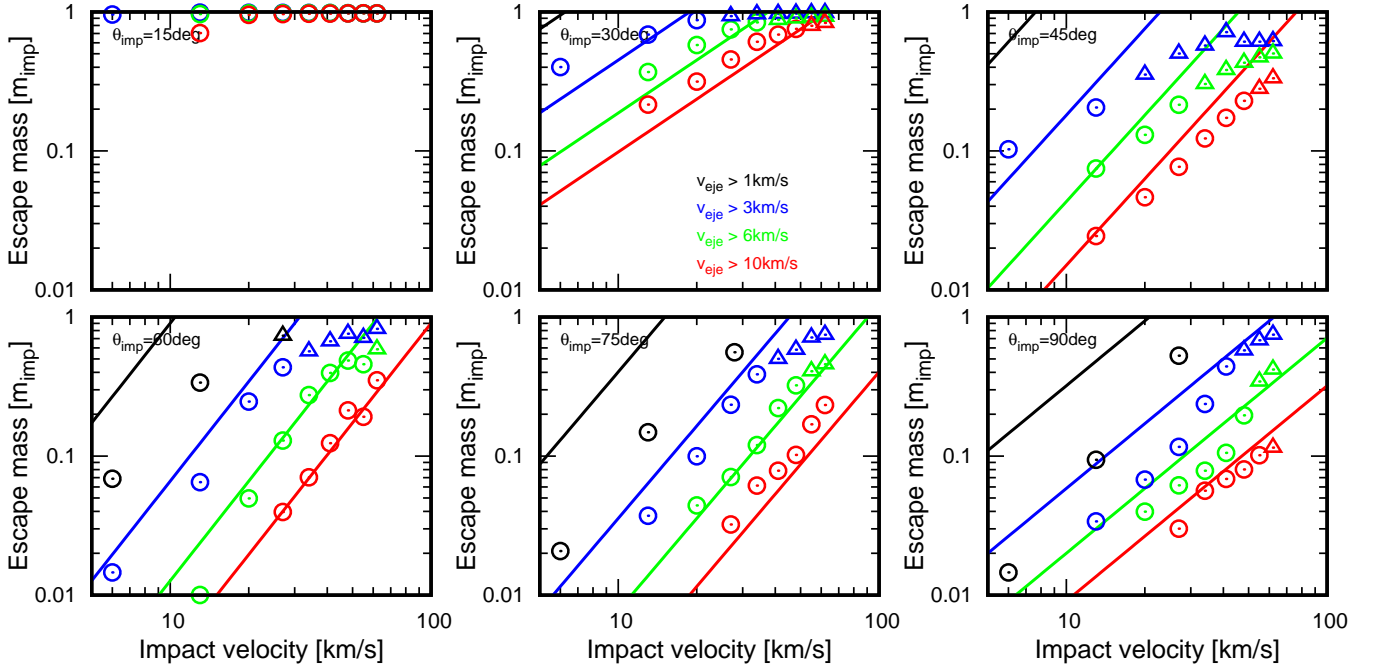}
\caption{Escape mass of the impactor material (in units of $m_{\rm imp}$) as a function of $v_{\rm imp}$ for different $\theta$. The points are the results of the SPH simulations. Triangles are used to indicate cases for which numerical simulations do not converge. Circles are employed when numerical simulations converge. Solid lines represent the new scaling law (Equation \ref{eq_HG21_impactor}) derived in this study. The black, blue, green, and red lines represent the cases of $v_{\rm esc}=1, 3, 6$, and $10$ km s$^{-1}$, respectively.}
\label{fig_impactor}
\end{figure*}

The derived escape mass of the impactor material is defined as $M_{\rm HG21,esc,imp}(>v_{\rm esc})$. Using the same arguments as Paper 1, we assume that $M_{\rm HG21,esc,imp}(>v_{\rm esc})$ follows a power-law function as follows:

%%%%%%%%%%%
% Equation 6
%%%%%%%%%%%
\begin{align}
\label{eq_HG21_impactor}
	 \frac{M_{\rm HG21,esc,imp}(>v_{\rm esc})}{m_{\rm imp}} = \nonumber \\
	 C_{\rm HG21,imp}(\theta) & \left( \frac{v_{\rm esc}}{v_{\rm imp}\sin(\theta)} \right)^{-3\mu_{\rm HG21,imp}(\theta)} ,
\end{align}

\noindent where $C_{\rm HG21,imp}(\theta)$ and $\mu_{\rm HG21,imp}(\theta)$ are coefficients and exponents that depend on the impact angle. Equation \ref{eq_HG21_impactor} is fit to the converged numerical results mainly at $v_{\rm esc}=10$ km s$^{-1}$ to obtain the coefficient and exponent at varying impact angles.

Figure \ref{fig_mu_C_impactor} shows the coefficients and exponents obtained from the numerical simulations. As in the case of the rocky bodies (Paper 1), we derived $\mu_{\rm HG21, imp}(\theta)$ and $C_{\rm HG21, imp}(\theta)$ (lines in Figure \ref{fig_mu_C_impactor}), respectively, as: 

%%%%%%%%%%%
% Equation 7&8
%%%%%%%%%%%
\begin{align}
\label{eq_mu_HG21}
	 & \mu_{\rm HG21,imp}(\theta) = a_{\rm imp}\theta^2 + b_{\rm imp}\theta + c_{\rm imp} \\
\label{eq_C_HG21}
	 & C_{\rm HG21,imp}(\theta) = \exp \left( d_{\rm imp}\theta^3 + e_{\rm imp}\theta^2 + f_{\rm imp}\theta + g_{\rm imp} \right) ,
\end{align}

\noindent where $a_{\rm imp}$, $b_{\rm imp}$, $c_{\rm imp}$, $d_{\rm imp}$, $e_{\rm imp}$, $f_{\rm imp}$, and $g_{\rm imp}$ are the fitted parameters, respectively (Table \ref{table_param}). Note that $\mu_{\rm HG21,imp}(\theta) = 0$ and $C_{\rm HG21,imp}(\theta) = 1$ for $\theta < 15$ degrees.

Figure \ref{fig_impactor} shows the escape mass of an impactor material with a speed greater than $v_{\rm esc}$. Points represent data obtained from the SPH simulations. Some of the simulations for high impact velocities and impact angles did not converge within a reasonable computational time (plotted by triangles), while converged parameters are plotted by circles.

The newly derived scaling law (Equation \ref{eq_HG21_impactor}) is plotted in Figure \ref{fig_impactor} by solid lines. The law generally agrees with the numerical results, especially for $\theta = 30 - 75 $ degrees. Note that impacts with $\theta=0$ and $90$ degrees do not statistically occur in planet formation, because the impact angle distribution is given by $\sin 2\theta$ \citep{Sho62}.

The accretion mass of the impactor material onto the target surface upon cratering impacts is given as:

%%%%%%%%%%%
% Equation 9
%%%%%%%%%%%
\begin{align}
\label{eq_accretion}
	\frac{M_{\rm HG21,acc,imp}(<v_{\rm esc})}{m_{\rm imp}} = \nonumber \\
	  1 - C_{\rm HG21,imp}(\theta) & \left( \frac{v_{\rm esc}}{v_{\rm imp}\sin \theta} \right)^{-3\mu_{\rm HG21,imp}(\theta)} .
\end{align}

Discarding factors of the order of unity difference, we found that our numerical results are compatible with \cite{Pie02} (case of icy impactors on Europa's icy surface with small porosity) and \cite{Yue13} (case of rocky impactors on the rocky surface of the Moon). Comparing the results of \cite{Ong10} (case of icy impactors on the rocky surface of the Moon) with our results, we found a notable difference in the impactor's accretion mass. Our numerical simulations are limited to the target and impactor having the same composition. We believe further research needs to be carried out to study the impact of various factors, such as materials of target and impactor, numerical methods used on the numerical results (e.g., 2D/3D, different EOSs, Eulerian/Lagrangian methods).

%%%%%%%%%%%%%%%%%%%%%%%%%%%%%%%%%%%%%%%%
% Section 5
%%%%%%%%%%%%%%%%%%%%%%%%%%%%%%%%%%%%%%%%
\section{Applications for planet formation}
\label{sec_application}

%%%%%%%%%%%%%%%%%%%%%%%%%%%%%%%%%%%%%%%%
\subsection{$\theta$-averaged escape and accretion masses} \label{sec_ave}

Cratering impacts of small bodies on a large planetary body occur much more frequently than collisions between similar-sized bodies. The statistical distribution of the impact angle follows $\sin 2\theta$ with a peak of $45$ degrees \citep{Sho62}. 

Using the newly derived scaling laws for the escape mass of icy target material (Equation \ref{eq_new_target}) and the accretion mass of icy impactor material (Equation \ref{eq_accretion}), one can derive the $\theta$-averaged scaling laws weighted over the $\sin 2\theta$ distribution (see Paper 1 for rocky bodies). These are as follows.

The $\theta$-averaged escape mass originating from the icy target as a function of impact velocity was derived as

%%%%%%%%%%%
% Equation 10&11
%%%%%%%%%%%
\begin{align}
\label{eq_theta_average_target_ice1}
	  \left< \frac{M_{\rm ice,esc,tar}(>v_{\rm esc})}{m_{\rm imp}} \right>_{\theta} =
	 & 0.013 \left( \frac{v_{\rm imp}}{v_{\rm esc}} \right)^{2.29} \\
	 & \left[ \mbox{Icy bodies;\ } v_{\rm imp} \lesssim 9 v_{\rm esc} \right]
	 \nonumber \\
\label{eq_theta_average_target_ice2}
	 \left< \frac{M_{\rm ice,esc,tar}(>v_{\rm esc})}{m_{\rm imp}} \right>_{\theta} =
	 & 0.051 \left( \frac{v_{\rm imp}}{v_{\rm esc}} \right)^{1.65} .\\
	 & \left[ \mbox{Icy bodies;\ } v_{\rm imp} \gtrsim 9 v_{\rm esc} \right]  \nonumber 
\end{align}

\noindent That for rocky bodies was (derived in Paper 1):

%%%%%%%%%%%
% Equation 12&13
%%%%%%%%%%%
\begin{align}
\label{eq_theta_average_target_rock1}
	  \left< \frac{M_{\rm rock,esc,tar}(>v_{\rm esc})}{m_{\rm imp}} \right>_{\theta} & =
	  0.02  \left( \frac{v_{\rm imp}}{v_{\rm esc}} \right)^{2.2} \\
	 & \left[ \mbox{Rocky bodies;\ } v_{\rm imp} \lesssim 12 v_{\rm esc} \right]
	 \nonumber \\
\label{eq_theta_average_target_rock2}
	 \left< \frac{M_{\rm rock,esc,tar}(>v_{\rm esc})}{m_{\rm imp}} \right>_{\theta} & =
	  0.076  \left( \frac{v_{\rm imp}}{v_{\rm esc}} \right)^{1.65} .\\
	 & \left[ \mbox{Rocky bodies;\ } v_{\rm imp} \gtrsim 12 v_{\rm esc} \right]  \nonumber 
\end{align}

\noindent In the above equations, one uses the smaller one of the two (Equations \ref{eq_theta_average_target_ice1} -- \ref{eq_theta_average_target_ice2} or Equations \ref{eq_theta_average_target_rock1} -- \ref{eq_theta_average_target_rock2}). Equations \ref{eq_theta_average_target_ice2} and \ref{eq_theta_average_target_rock2} are obtained by averaging the point-source scaling laws\footnote{From the point-source assumptions (Equation \ref{eq_HH11}), the $\theta$-averaged power-law dependences are $\propto (v_{\rm imp}/v_{\rm esc} )^{3\mu_{\rm HH11}} = (v_{\rm imp}/v_{\rm esc} )^{1.65}$ for both rocky and icy bodies. The $\theta$-averaged coefficients (see Equation \ref{eq_HH11}) are $<C_{\rm HH11}>_{\rm \theta} = C_{\rm HH11} \int_0^{\pi/2} (\sin \theta)^{3\mu_{\rm HH11}} \sin 2 \theta d \theta \simeq 0.051$ for icy bodies and $\simeq 0.076$ for rocky bodies, respectively.}.

The $\theta$-averaged accretion mass originating from the icy impactor as a function of impact velocity is

%%%%%%%%%%%
% Equation 14
%%%%%%%%%%%
\begin{align}
\label{eq_theta_average_accretion_ice}
	 \left< \frac{M_{\rm ice,acc,imp}(<v_{\rm esc})}{m_{\rm imp}} \right>_{\theta} = 
	  0.88 - & 0.024  \left( \frac{v_{\rm imp}}{v_{\rm esc}} \right)^{1.66}. \nonumber \\
	 & \left[ \mbox{Icy bodies} \right]
\end{align}

\noindent That for rocky bodies was (derived in Paper 1):

%%%%%%%%%%%
% Equation 15
%%%%%%%%%%%
\begin{align}
\label{eq_theta_average_accretion_rock}
	 \left< \frac{M_{\rm rock,acc,imp}(<v_{\rm esc})}{m_{\rm imp}} \right>_{\theta} = 
	 0.85 & - 0.071  \left( \frac{v_{\rm imp}}{v_{\rm esc}} \right)^{0.88}. \nonumber \\
	 & \left[ \mbox{Rocky bodies} \right]
\end{align}

These $\theta$-averaged scaling laws are shown in Figure \ref{fig_rocky_icy_bodies}. In Section \ref{sec_bodies}, the typical outcome of a cratering impact at a variety of planetary bodies in the solar system (see Table \ref{table_planet}) is discussed.

%%%%%%%%%%%%%%%%%%%%%%%%%%%%%%%%%%%%%%%%
\subsection{Critical target size where strength needs to be included} \label{sec_boundary}
Small impactors are attracted by the gravity of large target bodies (i.e., cratering impacts) and impact velocity becomes greater than the escape velocity of the target ($v_{\rm imp} \gtrsim v_{\rm esc}$). Our numerical simulations neglected material strength. In this subsection, using a simple but a well-defined analytical argument, we estimate the critical target size below which material strength affects the determination of ejecta having velocity greater than the escape velocity of the target (i.e., ejecta of $v_{\rm eje} \geq v_{\rm esc}$).

Using the Rankine-Hugoniot relations, the shock pressure $P_{\rm shock}$ during the cratering impact is given by
%
%%%%%%%%%%%
% Equation 16
%%%%%%%%%%%
\begin{align}
\label{eq_Pshock}
	P_{\rm shock} = \rho_{\rm 0} \left( C_{\rm 0} + s u_{\rm p} \right) u_{\rm p},
\end{align}
where $\rho_{\rm 0}$ is material density before the shock compression, $C_{\rm 0}$ is bulk sound velocity \citep[$C_{\rm 0} \simeq 1000-3000$ km s$^{-1}$ for rocky and icy materials, respectively;][]{Mel89}, $u_{\rm p}$ is a particle velocity ($u_{\rm p}$ is related to shock velocity as $u_{\rm s}= C_{\rm 0} +s u_{\rm p}$), and $s \equiv du_{\rm s}/du_{\rm p}$ is a constant \citep[$s \simeq 1-2$;][]{Mel89}.

For $P_{\rm shock} > P_{\rm HEL}$ where $P_{\rm HEL}$ is the Hugoniot elastic limit (HEL), a critical pressure above which fluid-like response is expected, the ejection process is expected to be unaffected by material strength. In this paper, the minimum ejection velocity of our primary interest is equal to escape velocity (i.e., $v_{\rm eje}=v_{\rm esc}$ and a larger ejection velocity accompanies a larger shock pressure). $u_{\rm eje}$ should be larger than $u_{\rm p}$ due to acceleration during adiabatic expansion. Approximately $u_{\rm eje} \sim 2u_{\rm p}$ \citep{Mel89}, which leads to $u_{\rm p}=v_{\rm esc}/2 = \sqrt{2G\pi \rho_{\rm 0}/3}R_{\rm tar}$ where $R_{\rm tar}$ is the target radius.

For $C_{\rm 0} > s u_{\rm p}$ (where $u_{\rm p} = v_{\rm esc}/2)$, we can approximate $ C_{\rm 0} + s u_{\rm p} \simeq C_{\rm 0}$ to determine the minimum shock pressure experienced by ejecta of $v_{\rm eje} \geq v_{\rm esc}$ during a cratering impact. This condition is rewritten as
%
%%%%%%%%%%%
% Equation 17
%%%%%%%%%%%
\begin{align}
\label{eq_Rcri}
	R_{\rm tar} & < C_{\rm 0}/(s\sqrt{2G\pi \rho_{\rm 0}/3}) \\
			 & < 2300 {\rm \, km} \times \\
			 & \hspace{3em} \left( \frac{C_{\rm 0}}{3000 {\rm \, m \, s^{-1}}} \right)   \left( \frac{s}{2} \right)^{-1}  \left( \frac{\rho_{\rm 0}}{3000 {\rm \, kg \, m^{-3}}} \right)^{-1/2} .
\end{align}

For a small sized target (Equation \ref{eq_Rcri}), we can determine the shock pressure for above condition by
%
%%%%%%%%%%%
% Equation 18
%%%%%%%%%%%
\begin{align}
\label{eq_Ppeak_app}
	P_{\rm shock} & = \rho_{\rm 0} \left( C_{\rm 0} + s u_{\rm p} \right) u_{\rm p} \simeq C_{\rm 0} \rho_{\rm 0} u_{\rm p} \\
	  & = \sqrt{2 \pi G/3} C_{\rm 0} \rho_{\rm 0}^{3/2}  R_{\rm tar} \\
	  & \simeq 600 {\rm \, MPa} \times \\ 
	  & \hspace{1.5em} \left( \frac{C_{\rm 0}}{3000 {\rm \, m \, s^{-1}}} \right)  \left( \frac{\rho_{\rm 0}}{3000 {\rm \, kg \, m^{-3}}} \right)^{3/2} \left( \frac{R_{\rm tar}}{100 {\rm \, km}} \right).
\end{align}
Typically $P_{\rm HEL} \simeq 2-5 {\rm \, GPa}$ for rocky bodies \citep{Sek08} and $P_{\rm HEL} \simeq 150-300 {\rm \, MPa}$ for icy bodies \citep{Ste03}. The critical size for $P_{\rm shock}=P_{\rm HEL}$ is readily obtained from Equation \ref{eq_Ppeak_app}. For icy bodies, $P_{\rm shock} > P_{\rm HEL}$ (i.e., strength is negligible) when $R_{\rm tar} \gtrsim 600$ km (with $P_{\rm HEL} \sim 300$ MPa, $C_{\rm 0} \sim 1300$ m s$^{-1}$, and $\rho_{\rm 0} \sim 1000$ kg m$^{-3}$). For rocky bodies, $P_{\rm shock} > P_{\rm HEL}$ when $R_{\rm tar} \gtrsim 1000$ km (with $P_{\rm HEL} \sim 5$ GPa, $C_{\rm 0} \sim 2600$ m s$^{-1}$, and $\rho_{\rm 0} \sim 2900$ kg m$^{-3}$).

We have ignored the effects of material strength in our study. However, further research needs to be carried out on the effects of material strength on the erosion mass and accretion mass.

%%%%%%%%%%%%%%%%%%%%%%%%%%%%%%%%%%%%%%%%
\subsection{Cratering impacts on rocky and icy bodies} \label{sec_bodies}

%%%%%%%%%%%%%%%%%%%%
% Table
%%%%%%%%%%%%%%%%%%%%
\begin{table*}[t]
\begin{center}
\begin{tabular}{|c|c|c|cl}
\hline
 Name & $v_{\rm imp}$ [km s$^{-1}$] & Description \\ 
 (rocky/icy) & ($v_{\rm imp}$ [$v_{\rm esc}$])  & (Reference) \\ \hline \hline
 Mercury  & $\sim 36$ km s$^{-1}$  & Instability phase \\ 
 (rocky)	 & ($\sim 8.5 v_{\rm esc}$; $v_{\rm esc} \simeq 4.25$ km s$^{-1}$) & \citep[e.g.,][]{Hyo21a}  \\  \hline
 Venus	 & $\sim 26 $ km s$^{-1}$  & Instability phase \\ 
 (rocky)	 & ($\sim 2.5 v_{\rm esc}$; $v_{\rm esc} \simeq 10.4$ km s$^{-1}$) & \citep[e.g.,][]{Moj19} \\  \hline
 Earth	 & $\sim 22 $ km s$^{-1}$  & Instability phase \\ 
 (rocky)	 & ($\sim 2.0 v_{\rm esc}$; $v_{\rm esc} \simeq 11.2$ km s$^{-1}$) &  \citep[e.g.,][]{Moj19} \\  \hline
 Moon     & $\sim 19 $ km s$^{-1}$  & Instability phase  \\ 
 (rocky)	 & ($\sim 8.0 v_{\rm esc}$; $v_{\rm esc} \simeq 2.4$ km s$^{-1}$) &  \citep[e.g.,][]{Bra20} \\  \hline
 Mars	 & $\sim 13 $ km s$^{-1}$  &  Instability phase\\ 
 (rocky) 	 & ($\sim 2.6 v_{\rm esc}$; $v_{\rm esc} \simeq 5.0$ km s$^{-1}$) &  \citep[e.g.,][]{Moj19} \\  \hline
 Ceres     & $\sim 5$ km s$^{-1}$  &  Today's population \\ 
 (rocky)	 & ($\sim 9.8 v_{\rm esc}$; $v_{\rm esc} \simeq 0.5$ km s$^{-1}$) &  \citep[e.g.,][]{Bot94} \\ \hline
 Vesta     & $\sim 5$ km s$^{-1}$  &  Today's population  \\ 
 (rocky)	& ($\sim 13.9 v_{\rm esc}$; $v_{\rm esc} \simeq 0.35$ km s$^{-1}$) &  \citep[e.g.,][]{Bot94} \\ \hline
 Rhea     & $\sim 15$ km s$^{-1}$  &   Instability phase \\ 
 (icy)       & ($\sim 23.6 v_{\rm esc}$; $v_{\rm esc} \simeq 0.64$ km s$^{-1}$) & \citep[e.g.,][]{Won21} \\  \hline
 Europa   & $\sim 24$ km s$^{-1}$  &   Instability phase \\ 
 (icy)       & ($\sim 12.0 v_{\rm esc}$; $v_{\rm esc} \simeq 2.0$ km s$^{-1}$) & \citep[e.g.,][]{Won21} \\  \hline
 Pluto      & $\sim 2$ km s$^{-1}$  & Today's population \\ 
 (icy)	 & ($\sim 1.7 v_{\rm esc}$; $v_{\rm esc} \simeq 1.2$ km s$^{-1}$) & \citep[e.g.,][]{Del13} \\  \hline
\end{tabular}

\caption{Examples of typical impact velocities for different rocky and icy bodies in different contexts and epochs. The strong gravity of giant planets attracts small bodies as high-speed impactors onto their moons. During the instability phase, the solar system likely experienced putative cataclysm leading to a rise in impact flux that characterized the surface morphology and composition of solar system bodies.}
\label{table_planet}
\end{center}
\end{table*}

%%%%%%%%%%%%%%%%%%%%
% Figure 8 with double column
%%%%%%%%%%%%%%%%%%%%
\begin{figure*}[ht!]
\centering \includegraphics[width=\textwidth]{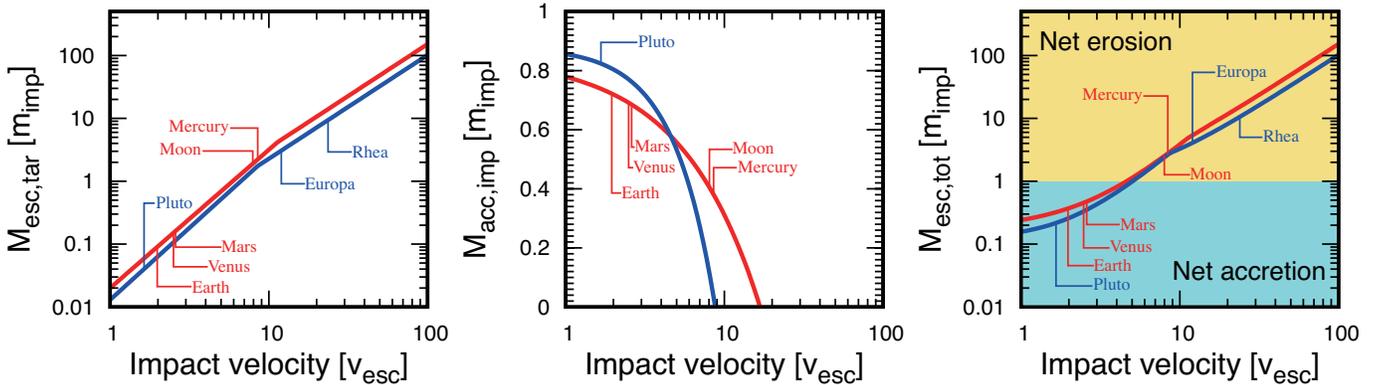}
\caption{$\theta$-averaged scaling laws for rocky (red lines) and icy (blue lines) bodies as a function of $v_{\rm imp}/v_{\rm esc}$. Left: the escape mass of the target material $M_{\rm esc,tar}/m_{\rm imp}$ (i.e., erosion mass of target; Equations \ref{eq_theta_average_target_ice1}--\ref{eq_theta_average_target_rock2}). Middle: the accretion mass of the impactor material onto the target surface $M_{\rm acc,imp}/m_{\rm imp}$ (Equations \ref{eq_theta_average_accretion_ice}--\ref{eq_theta_average_accretion_rock}). Right: total escape mass ($M_{\rm esc,tot} = M_{\rm esc,tar} + M_{\rm esc,imp}$ in a unit of $m_{\rm imp}$; $M_{\rm esc,tot} > 1$ indicates a net erosion and vice versa for a net accretion). Examples of typical outcomes of cratering impacts on different solar system bodies under different contexts are indicated by the labels in each panels (see Table \ref{table_planet} for the assumed impact conditions).}
\label{fig_rocky_icy_bodies}
\end{figure*}

In our solar system, a variety of orbits exist for planetary bodies including small bodies (asteroids and comets). For example, the orbital semi-major axis ranges from $a \sim 0.4$ to $40$ au for Mercury and Pluto, respectively. Cratering impacts inevitably occur in different contexts and epochs for a typical impact velocity range of $v_{\rm imp} \sim 1 - 100 v_{\rm esc}$ (Table \ref{table_planet}). High-speed cratering impacts would occur during the instability phase \citep{Moj19,Bra20}; these include the "Nice model" \citep[e.g.,][]{Gom05,Tsi05}, the "Grand-tack" hypothesis \citep[e.g.,][]{Wal11}, and the "early instability" scenario \citep[e.g.,][]{Cle18}. Giant impacts, such as those that formed the Moon \citep{Bot15} and/or the Martian moons \citep{Hyo18}, would distribute impact debris throughout the inner solar system, and high-speed collisions between the debris and asteroids (and/or planets) may correspondingly occur ($v_{\rm imp} \gg 5$ km s$^{-1}$).

Secular collisional evolution over billions of years and/or a rise of impact flux during the instability phase are expected to characterize the surface morphology and composition of solar system bodies. Figure \ref{fig_rocky_icy_bodies} illustrates examples of typical outcomes of cratering impacts that would potentially occur on different planetary bodies.

Comparing rocky and icy bodies, rocky target bodies are, in general, more eroded than icy target bodies (left panel in Figure \ref{fig_rocky_icy_bodies}), while less icy impactor's material is accreted on icy target surface than the case of rocky bodies for $v_{\rm imp} \gtrsim 4 v_{\rm esc}$ (middle panel in Figure \ref{fig_rocky_icy_bodies}).

Cratering impacts on rocky terrestrial planets, except Mercury, generally lead to a net mass increase as their $v_{\rm imp}/v_{\rm esc}$ ratios are typically $v_{\rm imp}/v_{\rm esc} \lesssim 5$ even in an instability phase. Mercury is the innermost planet in the solar system, leading to the highest orbital Keplerian velocity. Together with its small size, Mercury, cratering impacts on average lead to a net erosion \citep[see also][]{Hyo21a}. The typical impact velocity on the Moon in an instability phase is similar to that on Earth, whereas its mass is $\sim 0.01$ Earth mass. This leads to $v_{\rm imp}/v_{\rm esc} \sim 8$, and cratering impacts are generally erosive for the Moon.

The largest asteroids are Ceres ($\sim 500$ km in radius) and Vesta ($\sim 260$ km in radius). The current average impact velocity among asteroids is $\sim 5$ km s$^{-1}$ \citep[e.g.,][]{Bot94}. Therefore, $v_{\rm imp}/v_{\rm esc} \sim 10$ and $v_{\rm imp}/v_{\rm esc} \sim 14$ for Ceres and Vesta, respectively. Such a high-velocity impact is expected to be erosive. However, we note that our scaling laws might not be applicable to these asteroids because we have ignored material strength in our study (Section \ref{sec_boundary}).

Moons around giant planets and trans-Neptunian objects, such as Rhea, Europa, and Pluto, are icy bodies. The strong gravity of giant planets attracts small bodies as high-speed impactors \citep[e.g., $v_{\rm imp}/v_{\rm esc} \sim 23.6$ for the largest Saturn's regular moon, Rhea;][]{Won21} and erosive cratering impacts on the moons occur. The typical impact velocity on Pluto is $v_{\rm imp}/v_{\rm esc} \sim 1.7$ \citep{Del13}, which leads to a net accretion of cratering impacts.

%%%%%%%%%%%%%%%%%%%%%%%%%%%%%%%%%%%%%%%%
% Section 6
%%%%%%%%%%%%%%%%%%%%%%%%%%%%%%%%%%%%%%%%
\section{Summary} \label{sec_summary}

Cratering impacts inevitably occur during planet formation in different contexts and epochs. Cratering impacts erode a portion of the target surface by ejecting materials with a speed larger than its escape velocity while implanting a fraction of the impactor's material onto the target surface \citep[e.g.,][and references therein]{Hou83,Hou11,Hyo20}.

As was done in the companion study for rocky bodies (Paper 1), SPH simulations of cratering impacts on icy bodies were performed. The main findings are as follows: 

\begin{itemize}
\setlength{\parskip}{0cm} 
\setlength{\itemsep}{0.2cm}

 \item[--] Updated scaling laws were derived that predict the escape mass of the icy target material whose speed exceeds the escape velocity of a target (i.e., the erosion mass of the target; Equation \ref{eq_new_target}). It was demonstrated that a widely known classical point-source scaling law largely overestimated the escape mass of icy bodies when $v_{\rm imp} \lesssim 9 v_{\rm esc}$ (the same occurs when $v_{\rm imp} \lesssim 12v_{\rm esc}$ for rocky bodies; Paper 1).
 
 \item[--] Correspondingly, previous studies that have adopted the point-source scaling laws for rocky and icy bodies in an inappropriate $v_{\rm imp}/v_{\rm esc}$ regime should be redone. The conclusions derived in such studies would overestimate the erosion mass upon cratering impacts.
 
 \item[--] We also derived a scaling law that predicts the accretion mass of an impactor's material upon cratering impacts of icy bodies (Equation \ref{eq_accretion}; see Paper 1 for rocky bodies). The $\theta$-averaged scaling law weighted over $\sin2\theta$ (Equation \ref{eq_theta_average_accretion_ice}) indicates that no impactor material accretes on the target surface for $v_{\rm imp}/v_{\rm esc} \gtrsim 9$.

 \item[--] Statistically speaking (the $\theta$-averaged over $\sin 2\theta$), rocky bodies are generally more easily eroded than icy bodies (Equations \ref{eq_theta_average_target_ice1}--\ref{eq_theta_average_target_rock2}). Net erosion for icy and rocky bodies takes place for $v_{\rm imp}/v_{\rm esc} \gtrsim 5$ (Figure \ref{fig_rocky_icy_bodies}).

\end{itemize}

Diverse cratering impacts have critical roles in the mass gain/loss and in characterizing the surface composition of planets and asteroids (Figure \ref{fig_rocky_icy_bodies}). Further studies on cratering impacts are warranted for diverse planetary bodies.\\

%%%%%%%%%%%%%%%%%%%%%
% Acknowledgment
%%%%%%%%%%%%%%%%%%%%%
%\begin{acknowledgments}

\noindent R.H. acknowledges the financial support of JSPS Grants-in-Aid (JP17J01269, 18K13600). H.G. acknowledges the financial support of MEXT KAKENHI Grant (JP17H06457), and JSPS Kakenhi Grant (19H00726). We would like to thank Editage (www.editage.com) for English language editing.

%\end{acknowledgments}

\newpage
\bibliography{Cratering}
\bibliographystyle{aasjournal}

\end{document}